# Determination of the Energy Eigenvalues of the Varshni-Hellmann Potential

N.Tazimi

Department of Physics, University of Kashan, Kashan, Iran

## Abstract

In this paper, we solve the bound state problem for Varshni-Hellmann potential via a useful technique. In our technique, we obtain the bound state solution of the Schrödinger equation for the Varshni-Hellmann potential via ansatz method. We obtain the energy eigenvalues and the corresponding eigen-functions. Also, the behavior of the energy spectra for both the ground and the excited state of the two body systems is illustrated graphically. The similarity of our results to the accurate numerical values is indicative of the efficiency of our technique.

## 1 Introduction

Exact solution of Schrödinger equation in the D-dimensional coordinates system has been a focus of study in miscellaneous works of quantum physics and quantum chemistry. The energy eigenvalues and wave function, which are capable of showing the behavior of a quantum mechanical system, can be obtained from the Schrödinger equation. The Schrödinger equation is a second-order differential equation used to solve quantum-mechanics problems. The exact and approximate solutions of the Schrödinger wave equation in non-relativistic quantum mechanics have many features because the wave functions and their equivalent eigenvalues provide a lot of information for the description of various quantum systems, including atomic structure theory, quantum chemistry, and quantum electrodynamics. Using the experimental proof of the Schr¨odinger wave equation, researchers are motivated to solve the radial Schr¨odinger equation via different analytical methods. Attempts have been made to solve the Schr¨odinger and Klein-Gordon equations through different potentials. For example, William et al. studied the Hulthen potential together with the Hellmann potential [3] and Hans Hellmann investigated the Schr¨odinger equation with a linear combination of the Coulomb and Yukawa potentials, which is known as the Hellmann potential [4]. Hellman potential has been applied to several branches of physics such as atomic physics, plasma physics, solid state physics, etc. [5-6] and it has been used in the study of electron nucleus [7] and electron ion [8].

---

[1] tazimi@kashanu.ac.ir

This study seeks to obtain the eigenvalues and wave function of the 3D Schr¨odinger equation through the sum of Varshni and Hellmann potentials. The paper is organized as follows: In Sec. 2, the exact solution of the Schr¨odinger equation for Hellmann-Varshni potential is derived and we obtain the analytical expressions for energy levels and the corresponding wave functions for n and l quantum numbers. In Sec.3, the results are discussed. In Sec. 4, summary and conclusion are presented.

## 2    Formulation of the Approach

Schr¨odinger equation has been solved exactly by using various potentials, and it has been employed in different atomic, molecular and nuclear fields. Schr¨odinger equation is a second-order differential equation which serves to solve quantum-mechanic problems. We have attempted to solve Schr¨odinger and Klein-Gordon equations by using different potentials for few-quark systems [9-13]. In this section we solve Schr¨odinger equation by using Hellmann potential. Hellmann potential is of the following form: [ 14, 15]

$$V(r) = -\frac{c}{r} + \frac{d}{r}e^{-\alpha r} \qquad (1)$$

where r is the inter-nuclear interval. c and d stand for the strong points of Coloumb and Yukawa potentials. Varshni potential is of the following form: [16]

$$V(r) = a - \frac{ab}{r}e^{-\alpha r} \qquad (2)$$

where a and b denote the strong points of Varshni potential. Varshni potential is a function of repulsive short-range potential energy, which has been studied in the formalism of Schrödinger equation and contributed greatly to chemical and nuclear physics [17,18]. In this article we study Schrödinger radial equation with a new proposed potential obtained from the sum of Varshni and Hellmann potential (VHP). The potential is:

$$V(r) = a + \frac{d - ab}{r} e^{-\alpha r} - \frac{c}{r} \tag{3}$$

In fig. 1, we show the VHP Potential variations in terms of different values of $\alpha$. We expand

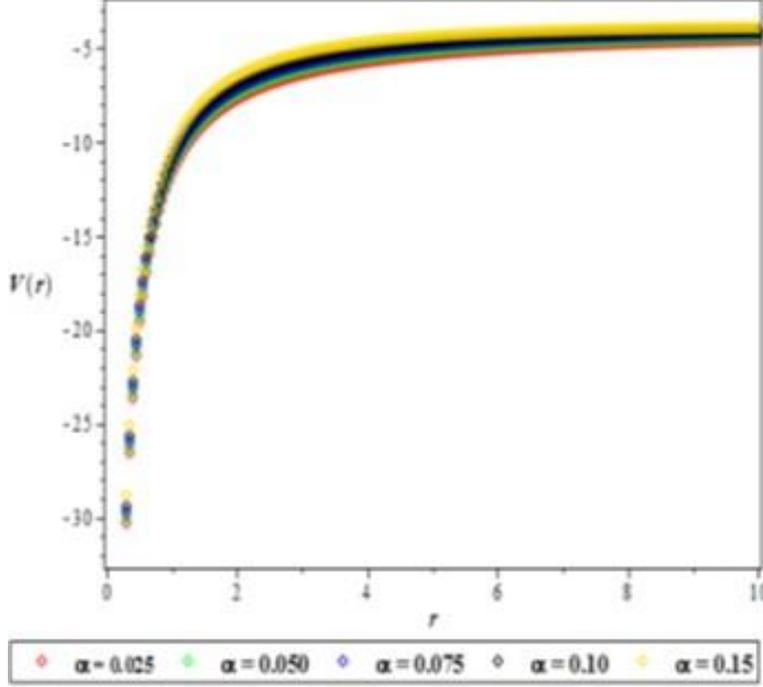

Figure 1: VHP Potential variations in terms of different values of $\alpha$

.

the exponential part of the potential:

$$V(r) = a + \frac{d - ab}{r}\left(1 - \alpha r + \frac{\alpha^2 r^2}{2} - \frac{\alpha^3 r^3}{6}\right) - \frac{c}{r} \tag{4}$$

And we write the potential in a simpler form:

$$V(r) = a + \frac{d - ab}{r} + (ab - d)\alpha - \frac{(ab - d)\alpha^2 r}{2} + \frac{(ab - d)\alpha^3 r^2}{6} - \frac{c}{r} \tag{5}$$

The systems Schrödinger equation is:

$$-\frac{\hbar^2}{2\mu} \frac{1}{r^2} \frac{\partial}{\partial r}\left(r^2 \frac{\partial \psi_{v,l}(r)}{\partial r}\right) + \left(V(r) - E_{v,l} + \frac{l(l+1)\hbar^2}{2\mu r^2}\right)\psi_{v,l}(r) = 0 \tag{6}$$

where $\mu$ is the reduced mass and $v$ and $l$ are group number and the orbital quantum number of one particle relative to another, respectively. By choosing $\psi(r) = \frac{1}{r}\phi(r)$, eq. 5 appears as:

$$\phi''(r) + \frac{2\mu}{\hbar^2}\left[E - V(r) - \frac{l(l+1)\hbar^2}{2\mu r^2}\right]\phi(r) = 0 \tag{7}$$

$\phi(r)$ can be derived from eq. 6. Assuming $\hbar = c = 1$ and $\phi(r) = f(r)exp[g(r)]$, we can pursue the calculations related to $\phi(r)$ function, and $f(r)$ and $g(r)$ functions are presented as:

$$jf_j(r) = \prod_{c=1}(r - \alpha_j) \qquad j = 1,2,... \tag{8}$$

where $n = 1,2,3,.$ and $f_0(r) = 1$, and the polynomial $g(r)$ is defined based on the type of potential. In this case, it is defined based on potential 3.

$$g(r) = \frac{-1}{2}Ar^2 + Br + \delta \ln r \tag{9}$$

From the above equations, we have:

$$\phi''(r) = \left[g''(r) + g'^2(r) + \frac{f''(r) + 2g'(r)f'(r)}{f(r)}\right]\phi(r) \tag{10}$$

And by introducing into eq.4:

$$\frac{-2\mu}{\hbar^2}\left[E - V(r) - \frac{l(l+1)\hbar^2}{2Mr^2}\right] = \left[g''(r) + g'^2(r) + \frac{f''(r) + 2g'(r)f'(r)}{f(r)}\right] \tag{11}$$

We expand the exponential part of the potential and rewrite the potential as:

$$V(r) = a + \frac{d - ab}{r}(1 - \frac{\alpha}{r} - \frac{\alpha^2 r^2}{2} - \frac{\alpha^3 r^3}{6}) - \frac{c}{r} \tag{12}$$

By introducing the potential quantity and the derivatives into eq. 10, we solve the equation for n=0 and angular momentum L, and the following equation is obtained:

$$-2\mu(E - a + \frac{ab - d}{r} - (ab - d)\alpha + \frac{ab\alpha^2 r}{2} + \frac{(d - ab)\alpha^3 r^2}{6} + \frac{c}{r}$$
$$-\frac{l(l+1)}{2\mu r^2}) = -A + B^2 - 2A\delta + \frac{2B\delta}{r} - 2ABr + A^2 r^2 + \frac{\delta^2}{r^2} - \frac{\delta}{r^2} \tag{13}$$

With a simple calculation and considering that the exponents of $r$ are linearly independent, it is possible to set the coefficients of different powers of $r$ equal to each other. In this case, the following relations are obtained between the potential coefficients and the energy can be obtained:

$$-2\mu(E - a + (d - ab)\alpha) = -A + B^2 - 2A\delta \quad , \quad -2M\left(\frac{ab-d}{r} + \frac{c}{r}\right) = \frac{2B\delta}{r}$$

$$-2\mu\left(\frac{(ab-d)\alpha^2 r}{2}\right) = -2ABr \quad , \quad -2\mu\left(-\frac{(ab-d)\alpha^3 r^2}{6}\right) = -A^2 r^2$$

$$l(l+1) = \frac{\delta^2}{r^2} - \frac{\delta}{r^2}$$

(14)

By solving the above equations, the special relation of the energy values for the state $n = 0$ is obtained as follows.

$$E_{0L} = a + (ab-d)\alpha - \frac{(l+1)\left(\frac{(ab-d)\alpha^2}{2}\right)}{2\mu(ab+c-d)}(2l+3) - \frac{2\mu(ab+c-d)^2}{4(l+1)^2} \tag{15}$$

$$\psi_{0L} = N_1 r^{L+1} exp\left(\frac{-1}{2}Ar^2 + Br\right) \tag{16}$$

The energy for the first excited state $n = 1$ and the angular momentum L is equal to:

$$E_{1L} = a + (ab-d)\alpha - \frac{(l+1)\left(\frac{ab\alpha^2}{2} - \frac{d\alpha^2}{2}\right)}{2\mu(ab+c-d)}(2l+5) - \frac{2\mu(ab+c-d)^2}{4(l+2)^2} \tag{17}$$

## 3  Numerical Results

The detailed analysis of the results in terms of various domains of parameters $a,b,c$ and $\alpha$ of the VHP potential reveals a few important facts concerning the application of the perturbed formalism. In the present study the discrete energy eigenvalues for the VHP potential have been calculated as functions of the strength $a,b,c$ and the screening parameter $\alpha$ of the VHP potential.

1-    For VHP potential the energy eigenvalues is given by Eq. (15) and Eq. (17). In Table 1, we show the energy eigenvalues for VHP potential (with $a = 1, b = −1, c = 4, d = −4, \sim = 2\mu = 1$) in terms of different values of $\alpha$. As alpha increases, the magnitude of the binding energy decreases. The energy values corresponding to the states $n = 2,3,..$ are also obtained by the same method. In this way, the Schr¨odinger equation was solved analytically and the eigenvalues of $E_{nL}$ were obtained . In Fig 2 and 3, we show the variation of energy as a function of $\alpha$ and $M$ for different $l$ by using results of table 1. As we can see, for $l > 0$ the energy increases with the increase of $\alpha$ and the energy decreases with the increase of $\mu$.

2-    We have shown the energy eigenvalue for Helman potential for $a = b = 0$ in Table 2 and compared our results with [20] and [21]. Hamzavi et al in [20] have obtained the approximate analytical solutions of the radial Schr¨odinger equation for the Hellmann potential By using the generalized parametric Nikiforov-Uvarov (NU) method. Ref. [21] a perturbative treatment for the bound states of the Hellman potential is used. As we have shown in Table 2, as the value of $\alpha$ increases, the correlation energy decreases.

3- We have shown energy eigenvalue for Varshni potential $a = b = -1, (\hbar = 2\mu = 1)$ in table 3 and compared our results with Ref. [19]. Ebomwonyi *et al*. have studied the Schrödinger equation for the Varshni potential function with two eigen-solution techniques such as the NU and the semi-classical WKB approximation methods in Ref. [19]. Our results are in good agreement with Ref. [19].

4- Table 4 shows numerical values of the binding energies of Yukawa potential ($a = b = c = 0$) in terms of the values of $M$ and $\alpha$. The results obtained are compared with those of [22,23]. Ref. [22] applies the asymptotic iteration method to solve the radial Schrödinger equation for the Yukawa type potentials. Accurate numerical solutions have been obtained for Schrödinger equation through a Yukawa potential in Ref. [23].

## 4 Conclusions

In this research, we analyzed the Schrödinger equation with Varshni-Hellman potential using the Ansatz method. We study the discrete energy eigenvalues for the Hellmann-Varshni potential have been calculated as functions of the screening parameter $\alpha$ of the Yukawa potential. We compared our findings with other theoretical formalisms. We found that the energy eigenvalues obtained using this method are in good agreement with other works in the literature. Therefore, Analytical solutions while opens a new window It can be used to provide valuable information about the dynamics of quantum mechanics in molecular and atomic physics.

## 5 Conflicts of Interest

The authors declare that they have no conflicts of interest.

## 6 Data Availability

The data used to support the findings of this study are included within the article.

## 7 Acknowledgments

The authors received no financial support for the publication of this article.

Table 1: Energy eigen value for VHP potential in terms of different values of $\alpha$ ($a = 1, b = -1, c = 4, d = -4, \hbar = 2\mu = 1$)

| State | $\alpha$ | E(ev) |
|---|---|---|
| 1s | 0.025 | -19.175401 |
|    | 0.050 | -19.101607 |
|    | 0.075 | -19.028616 |
| 2s | 0.025 | -4.058816 |
|    | 0.050 | -4.041517 |
|    | 0.075 | -4.034352 |
| 2p | 0.025 | -4.028798 |
|    | 0.050 | -4.021861 |
|    | 0.075 | -3.952986 |
| 3s | 0.025 | -1.247048 |
|    | 0.050 | -1.234861 |
|    | 0.075 | -1.221445 |
| 3p | 0.025 | -1.232069 |
|    | 0.050 | -1.227473 |
|    | 0.075 | -1.143767 |
| 3d | 0.025 | -1.215805 |
|    | 0.050 | -1.172321 |
|    | 0.075 | -1.139522 |

Table 2: Energy eigenvalue for Helman potential $a = b = 0, c = 2, d = -1, (\hbar = 2\mu = 1)$

| state | $\alpha$ | E (ev) | Ref. [21] | Ref. [20] |
|---|---|---|---|---|

| state | $\alpha$ | | | |
|---|---|---|---|---|
| 1s | 0.001 | - 2.238 00 | - 2.249 00 | - 2.24898 |
|    | 0.005 | - 2.244 01 | - 2.245 01 | - 2.24499 |
|    | 0.01  | - 2.2413   | - 2.240 05 | - 2.24003 |
| 2s | 0.001 | - 0.5606   | - 0.561 50 | - 0.56150 |
|    | 0.005 | - 0.5569   | - 0.557 55 | - 0.55754 |
|    | 0.01  | - 0.55198  | - 0.552 69 | - 0.55269 |
| 2p | 0.001 | - 0.5601   | - 0.561 50 | - 0.56150 |
|    | 0.005 | - 0.5562   | - 0.557 54 | - 0.55754 |
|    | 0.01  | - 0.5516   | -0.552 66  | - 0.55266 |
| 3s | 0.001 | - 0.2381   | - 0.249 00 | - 0.24900 |
|    | 0.005 | - 0.24378  | - 0.245 11 | - 0.245 11 |
|    | 0.01  | - 0.2399   | - 0.240 43 | - 0.24043 |
| 3p | 0.001 | - 0.2478   | - 0.249 00 | - 0.24900 |
|    | 0.005 | - 0.2436   | - 0.245 10 | - 0.24510 |
|    | 0.01  | - 0.2390   | - 0.240 40 | - 0.24040 |
| 3d | 0.001 | - 0.2468   | - 0.249 00 | - 0.24900 |
|    | 0.005 | - 0.2446   | - 0.245 08 | - 0.24508 |
|    | 0.01  | - 0.2389   | - 0.240 34 | - 0.24034 |

Table 3: Energy eigenvalue for Varshni potential $a = b = -1, (\sim = 2\mu = 1)$

| state | $\alpha$ | E (ev) | [19] |
|---|---|---|---|
| 1s | 0.001 | - 1.249001 | - |
|    | 0.050 | - 1.203750 | - |
|    | 0.100 | - 1.165000 | - |
| 2s | 0.001 | - 1.0615025 | - |
|    | 0.050 | - 1.0187500 | - |
|    | 0.100 | - 0.9875000 | - |
| 2p | 0.001 | - 1.026784 | - 1.061750 |
|  | 0.050 | - 0.995277 | - 1.0256250 |
|  | 0.100 | - 0.997777 | - 0.990000 |
| 3s | 0.001 | - 1.026781 | - |
|  | 0.05 | - 0.9865277 | - |
|  | 0.1 | - 0.9627777 | - |
| 3p | 0.001 | - 1.014634 | - |
|  | 0.050 | - 0.988125 | - |
|  | 0.100 | - 1.005625 | - |
| 3d | 0.001 | - 1.0090165 | - 1.026944 |
|  | 0.05 | - 1.001250 | - 0.986736 |
|  | 0.1 | - 1.075000 | - 0.946944 |
| 4s | 0.001 | - 1.014629 | - |
|  | 0.050 | - 0.976875 | - |
|  | 0.100 | - 0.960625 | - |
| 4p | 0.001 | - 1.009011 | - 1.01506 |
|  | 0.05 | - 0.987500 | - 0.99515 |
|  | 0.1 | - 1.020000 | - 0.990000 |
| 4d | 0.01 | - 0.992075 | - 1.01493 |
|  | 0.050 | - 0.991805 | - 0.98515 |
|  | 0.100 | - 1.088055 | - 0.96250 |
| 4f | 0.01 | - 1.004132 | - 1.01475 |
|  | 0.050 | - 1.030102 | - 0.97250 |
|  | 0.100 | - 1.205102 | - 0.97250 |

Table 4: Energy eigenvalue for Yukawa potential $d = \sqrt{2}, \bar{\alpha} = gd (\sim = \mu = 1)$

| state | $g$ | E (ev) | Ref. [23] | Ref. [22] |
|---|---|---|---|---|
| 1s | 0.002 | - 1.00133 | - 0.99600 | - 0.99600 |
|     | 0.005 | - 1.00518 | - 0.99004 | - 0.99003  |
|     | 0.010 | - 0.9901  | - 0.98015 | - 0.98014  |
|     | 0.020 | - 0.9519  | - 0.96059 | - 0.96059  |
| 2s  | 0.002 | - 0.2378  | - 0.24602 | - 0.24602  |
|     | 0.005 | - 0.2396  | - 0.24015 | - 0.24014  |
|     | 0.010 | - 0.2276  | - 0.23059 | - 0.23058  |
|     | 0.020 | - 0.2105  | - 0.21230 | - 0.21229  |
| 2p  | 0.002 | - 0.2455  | - 0.24602 | - 0.24601  |
|     | 0.005 | - 0.2391  | - 0.24012 | - 0.24012  |
|     | 0.010 | - 0.22860 | - 0.23049 | - 0.23049  |
|     | 0.020 | - 0.21101 | - 0.21192 | - 0.21192  |
| 3p  | 0.002 | - 0.1067  | - 0.10716 | - 0.10716  |
|     | 0.005 | - 0.1009  | - 0.10142 | - 0.10141  |
|     | 0.010 | - 0.09076 | - 0.09231 | - 0.09230  |
|     | 0.020 | - 0.07489 | - 0.07570 | - 0.07570  |
| 3d  | 0.002 | - 0.1067  | - 0.10715 | - 0.10715  |
|     | 0.005 | - 0.1001  | - 0.10140 | - 0.10136  |
|     | 0.010 | - 0.0916  | - 0.09212 | - 0.09212  |
|     | 0.020 | - 0.0747  | - 0.07502 | - 0.07503  |

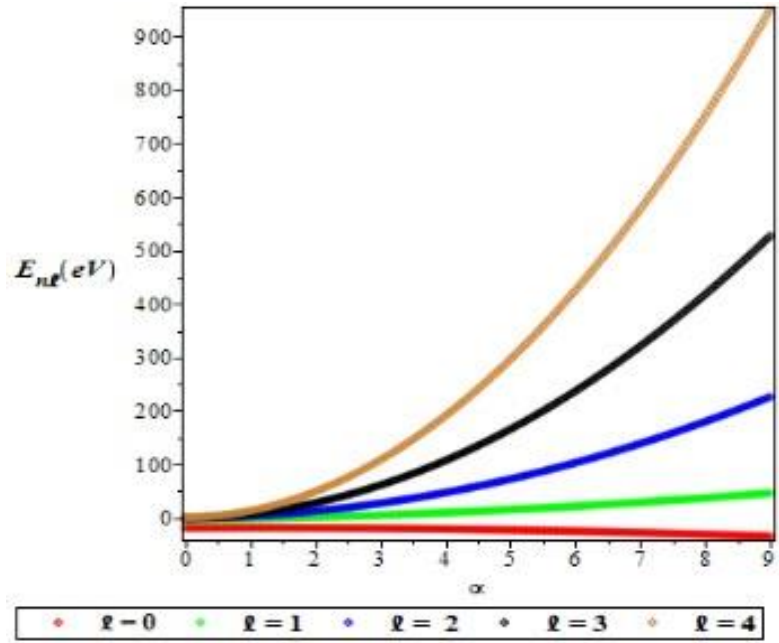

Figure 2: E in terms of $\alpha$ for different $l$

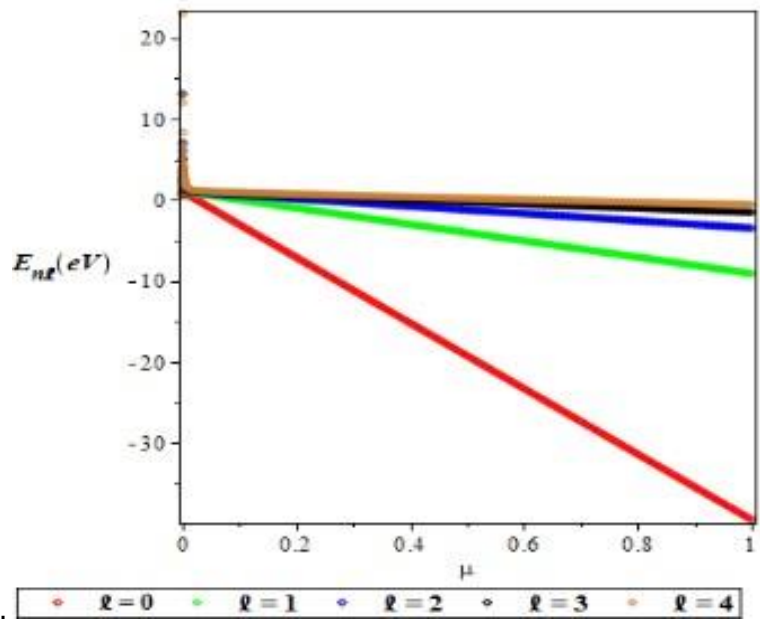

Figure 3: E in terms of of $\mu$ for different $l$